\documentclass{article}

\usepackage[english]{babel}
\usepackage{soul}

\usepackage{natbib}
\setcitestyle{authoryear,open={(},close={)}} 

\usepackage[letterpaper,top=2cm,bottom=2cm,left=3cm,right=3cm,marginparwidth=1.75cm]{geometry}

\usepackage{caption}
\usepackage{subcaption}
\usepackage{float}

\usepackage{amsmath}
\usepackage{graphicx}
\usepackage[colorlinks=true, allcolors=blue]{hyperref}

\usepackage{setspace}
\usepackage{parskip}
\usepackage{authblk}

\title{A Bayesian approach for checking and estimating block designs in agricultural experiments}
\author[1]{Josefina Lacasa}
\author[2]{Ignacio Ciampitti}
\author[1]{Trevor Hefley}
\affil[1]{Department of Statistics, Kansas State University}
\affil[2]{Department of Agronomy, Purdue University}

\date{} 

\begin{document}
\maketitle

\begin{abstract}
    Spatial blocking is a common technique used in designed agricultural experiments. 
    Spatial blocking, however, requires important assumptions about the spatial dependence patterns in the field. 
    Statistical models of blocked agricultural experiments often show spatial structure in the residuals, suggesting that the original model may benefit from additional assumptions regarding the spatial dependence patterns present in the field. 
    We propose a Bayesian model for checking block membership and estimating spatial patterns missing in the original block design. 
    Our model is particularly motivated by cases with mis-specified blocks.
    To accomplish this, we embed a regression tree within a Bayesian statistical model to estimate spatial effects with a block-like spatial structure. 
    We illustrate the benefits of our model using simulated data mimicking four typical scenarios we have encountered in our applied experience in agricultural research and using two datasets generated from designed agricultural experiments.
    We found that our model may serve two purposes: validating the originally designed block layout, or mis-specified block-like spatial structures. 
    Thus, our model provides researchers with a novel tool for analyzing data from designed agricultural experiments.
        
\end{abstract}

\section{Introduction}
\label{makereference1.1}

Spatial blocking is a widely used technique in designed agricultural experiments. 
Spatial blocks were introduced to agricultural experiments in the early 20th century as a surrogate for spatial effects. The notion of blocking is described as `local control' in \citet{fischer1925} (see also \citealp{Verdooren2020}, to learn about the history of designed experiments in agriculture). 
The concept of spatial blocks consists of partitioning an agricultural field into disjoint areas with similar characteristics such as soil fertility or topography. 
The notion of spatial blocks is present in several types of experiment designs, such as randomized complete block (RCBD), incomplete block, split-plot, strip plot, among others. 
The purported benefit of blocked experiment designs is an increased precision in the estimated treatment effects as compared to those from completely randomized designs \citep{milliken2009analysis}. 
This, in turn, enables more precise estimates of the treatment contrasts, which is usually the primary goal of agricultural experiments. 
Blocked designs, however, make at least two important assumptions: 1) the spatial blocks can be clearly delineated within a field by an agricultural researcher, and 2) the spatial blocks represent groups of similar experimental units (e.g., plots in the field; \citealt{milliken2009analysis}, page 74; \citealt{Mead2012}, page 9). 
	
Statistical analyses of data generated by blocked agricultural experiments often show spatial structure in the residuals (i.e., spatial correlation, \citealt{Brownie1993,Pugh2019a,StroupGLMM2024}), suggesting that one or both of the assumptions mentioned above were not met. 
Spatially correlated residuals may arise when blocks are forced to be large in area. 
For example, testing a large number (e.g., $\geq$ 100 genotypes, \citealp{Brownie1997,Piepho2006}) of treatments in an RCBD, or using farming equipment or technology are common situations that force larger than desired block sizes \citep{Borges2019}. 
An RCBD requires that each block is large enough to fit all treatments or equipment, and thus both cases require blocks that are larger in area than what might be available. 
However, in practice, RCBD experiments with many treatments result in larger contrived blocks, which are prone to violate the assumption of approximately similar plots within blocks \citep{Gusmao1986,StroupGLMM2024}. 
As a consequence, experiments with larger blocks often present spatial structure in the residuals. 
Eventually, unaddressed spatial effects can result in deleterious effects on the precision of the estimated treatment contrasts and thus misleading inference \citep{Stroup1994, Stroup2002}. 
Although in theory the estimator remains unbiased, the randomization scheme may place some treatments in more favorable areas and result in misleading treatment estimates.
For example, \citet{Stroup2002} showed that unaddressed spatial effects in a wheat trial with substantial spatial effects resulted in a misleading ranking of genotypes. 
In \citet{Stroup2002}, some genotypes were randomly placed in the worst areas of the field while other genotypes were randomly placed in the best areas of the field. 
Said spatial patterns (i.e., ``best" and ``worst" areas in the field) were not captured by the designed spatial blocks. 
\citet{Stroup2002} shows that said unaddressed spatial patterns resulted in a misleading ranking of the genotypes.

Methods borrowed from spatial statistics were introduced to improve statistical inference in the presence of spatial processes that were not considered in the design or cannot be represented by spatial blocks.  
The first analyses were performed using nearest neighbour analyses \citep{Bartlett1978, papadakis1937methode, Wilkinson1983} followed by trend analyses \citep{Federer1954}, smooth spatial splines (\citealt{gbur2020analysis}, chapter 6) and spatial random effects \citep{Federer1997, RICHTER2015}. 
The general advice for modeling data generated by designed agricultural experiments in the presence of spatial variability is to include both the elements of the design (e.g., spatial blocks; \citealt{Frey2024}) and the spatial position using one of the methods mentioned above \citep{Borges2019,Federer1997,Stroup2002}. 
It is worth pointing out that said spatial methods mentioned above typically assume smooth spatial processes.
Interestingly, in our applied experience we have often encountered non-traditional spatial dependence (e.g., \citealt{Mohankumar2022}) which may be generated by previous field management like tillage practices, crop species, and fertilization. 
Said management practices are usually applied on a field in a block-like manner and may justify the need for special modeling assumptions beyond traditional spatial dependence patterns.

In our applied experience in agricultural research we have found that elements from traditional experiment design (i.e., the notion of spatial blocks) are practical for most field experiments, but that spatial blocks may be mis-specified.  
Thus, our main goal is to introduce a hybrid model that can estimate block-like spatial effects that are ancillary to the original experiment design. 
Our hybrid model will enable agricultural researchers to check their original block layout and potentially model the potentially mis-specified blocks. 
More precisely, we introduce a hybrid model that combines model components from traditional experiment design (e.g., estimation of treatment and designed block effects) with regression trees (RT) to estimate spatial effects with block-like spatial structure, which enables inference about the parameters and adjusting the knowledge about block features (e.g., block boundaries, block membership, etc.). 
We illustrate the benefits of our hybrid model using simulated data mimicking four typical scenarios we have encountered in our applied experience in agricultural research and using two real datasets generated from designed agricultural experiments. 

\section{Model specifications}
\label{makereference1.2}

We propose a hybrid model that combines model components from traditional experiment design and can identify disjoint areas within an agricultural field that contain similar plots.
We conceptualize our model under the generalized linear model (GLM) framework (e.g., \citealp{stroup2012generalized, StroupGLMM2024})
using notation common for Bayesian models. 
Under this framework, a model for a blocked experiment can be generally described as
\begin{equation}\label{eq:1}
    y_i \sim [y_i | \mu_i, \psi]
\end{equation}
\begin{equation}\label{eq:2}
    g(\mu_i) = {\mathbf{x}_i}' \boldsymbol{\beta} + \eta_i,
\end{equation}
where $y_i$ is the $i$th observation unique to each plot ($i = 1, 2, ..., n$) and arises from a distribution of the exponential family, denoted by $[\cdot]$. 
The $\mu_i$ and $\psi$ are the expected value and dispersion of $[\cdot]$, respectively. 
The function $g(\cdot)$ is the link function (e.g., probit or logit). The inverse of the link function, $g^{-1}(\cdot)$, transforms the linear predictor ${\mathbf{x}_i}' \boldsymbol{\beta} + \eta_i$ to the same support as $\mu_i$. 
In the linear predictor ${\mathbf{x}_i} \equiv (1, x_{1,i}, x_{2,i}..., x_{p,i})'$, which indicates the treatment (of a total of $p$ treatments) applied to each plot and $\boldsymbol{\beta} \equiv (\beta_0, \beta_1, ..., \beta_p)'$ contains the treatment effects. The $\eta_i$ denotes the block effect (or spatial effect) of the $i$th plot. 
Estimating the treatment effects in $\boldsymbol{\beta}$ is usually the goal in agricultural experiments, while $\eta_i$ is generally regarded as a nuisance parameter.

The notion of spatial blocks is present in several types of experiment designs, like the RCBD and the incomplete block design which specify the block effects as $\eta_i = {\mathbf{z}_i}' \mathbf{u}$, where $\mathbf{z}_i$ indicates the assumed block membership (e.g., $\mathbf{z}_i = (0, 0, 1, 0)'$), and $\mathbf{u} \equiv (u_1, u_2, ..., u_K)'$ contains the block effects for $K$ blocks (in this case, $K=4$).  
There are currently two mainstream approaches to specify the block effects.
Both approaches assume $\mathbf{z}_i$ is known. 
The first approach makes no additional assumptions about $\mathbf{u}$ and is generally known as the ``fixed block effect", while the second approach models $\mathbf{u}$ as a random variable arising from a probability density function and is known as the ``random block effect". 
The most common probability density function specified for the random block effect is $\mathbf{u} \sim \text{N}(\boldsymbol{0}, \sigma^2_{u}\mathbf{I})$, which is also commonly expressed as $\boldsymbol{\eta} \sim \text{N}(0, \sigma^2_{u} \mathbf{Z} \mathbf{Z}')$, where $\boldsymbol{\eta} \equiv (\eta_1, ..., \eta_n)'$, $\mathbf{Z} \equiv (\mathbf{z}_1', \mathbf{z}_2', ..., \mathbf{z}_K')$, and $\sigma^2_{u} \mathbf{Z} \mathbf{Z}'$ is known as the compound symmetry variance-covariance matrix. 

Statisticians have provided a limited number of modern ways to specify the spatial effects $\boldsymbol{\eta}$ for data generated by designed agricultural experiments. 
The original approach involved including spatial blocks as a fixed effect \citep{fischer1925}.
With time, it became evident that the spatial blocks specified in the design are not always representing the true spatial processes of modern designed experiments. 
The supposedly known $\mathbf{Z}$ matrix does not always appear to correctly represent the spatial effects present in the field, as mentioned before. 
Thus, methods borrowed from spatial statistics were introduced to aid analyses where the classical approach yielded residuals with spatial structure \citep{Federer1954,papadakis1937methode, Wilkinson1983}. 
Said spatial methods are now considered standard solutions to aid agricultural researchers salvaging their studies in situations where the blocks do not capture all spatial processes (e.g., chapter 18 in \citealp{StroupGLMM2024}, chapter 6 in \citealp{gbur2020analysis}). 
Beyond these spatial corrections, further modern developments in statistical modeling appear absent in agriculture (e.g., Dirichlet processes, regression trees) or are underutilized (e.g., non-Gaussian random effects).

The available modeling options to specify the spatial effects beyond the designed blocks typically assume smooth spatial processes. 
However, non-traditional, block-like spatial effects may arise in field settings (e.g., due to previous field management).
Said block-like spatial effects may be considered mis-specified spatial blocks. 
In what follows, we describe our proposed model that is particularly motivated by such cases where blocks-like spatial effects exist, but are mis-specified.
Our approach expands the modeling options available to agricultural researchers, and may serve two purposes: validating the originally designed block layout, or estimating spatial effects with a block-like spatial structure.

Recall that the general advice to model data generated by designed experiments under the presence of spatial variability is to include the design components (e.g., spatial blocks) and a spatial effects component (e.g., spatial splines). 
In other words, said approach models spatial effects using two different model components.  
Likewise, we use the GLM framework in (\ref{eq:1}) and expand $\eta_i$ in (\ref{eq:2}) into a multi-resolution spatial component. 
Each $\eta_i$ may be dissected into the sum of spatial block components with different spatial resolutions. 
In this case, we consider only 2 components. 
The resolution of the first component is fixed at the resolution of the designed blocks, and the resolution of the second component is free, resulting in the estimation of spatial effects with block-like structure:
\begin{equation}\label{eq:3}
    \eta_i = \sum_{r=1}^{2} \mathbf{z}_{r_i}' \mathbf{u}_r,
\end{equation}
where $r=1,2$ is the resolution index, $\mathbf{z}_{r_i}'$ represents the block membership for the $i$th observation and resolution $r$, and $\mathbf{u}_r$ represents the spatial effects with a block-like spatial structure for resolution $r$. 
We consider that (\ref{eq:3}) with $r=1$ contains the information about the experiment design in ${\mathbf{z}_{1i}'}$ (i.e., designed block membership) and the block effects in $\mathbf{u}_1$.
With the designed blocks included in the model, we preserve all elements from the original design and avoid issues due to excluding the units where the randomization took place (i.e., blocks). 
Excluding the units where the randomization took place may lead to type I error inflation \citep{Frey2024} and produce misleading results \citep{Federer1997,Gilmour2000}.
In this case, we model the designed blocks as fixed and set a sum-to-zero restriction for the block effects, $\sum_{k=1}^{K} u_{1_k} = 0$, which is common among designed experiments.  
Furthermore, we consider that (\ref{eq:3}) with $r=2$ represents other spatial processes that are not accounted by the designed spatial blocks but have a spatial structure mimicking blocks. 
At $r=2$, our model estimates the effect $\mathbf{u}_2$ and the layout ${\mathbf{z}_{2_i}'}$ of said spatial effects with a block-like spatial structure. 

Heuristically, estimating spatial effects with a block-like spatial structure involves using an RT that partitions an agricultural field into groups of similar, contiguous, plots. 
Again, heuristically, the RT serves to estimate both $\mathbf{Z}_2$ (i.e., block membership in a different resolution than the experiment design) and $\mathbf{u}_2$.
We accomplish this by employing a Bayesian approach that relies on the models and Markov chain Monte Carlo algorithms proposed by \citet{Shaby2012}. 
The results provided by \citet{Shaby2012} are appealing to agricultural researchers (and statisticians) because they provide an easy-to-use approach that enables software from traditional machine learning algorithms (available in software such as R and Python) to be embedded within Bayesian GLMs.
Beyond the expertise required to program Bayesian GLMs from scratch (e.g., at the level of \citealp{hooten2019bringing}), implementing the \citet{Shaby2012} approach requires virtually no additional skill.
This is attractive for our study because, in our experience, implementation of Bayesian RTs (e.g., \citealp{Chipman1998}) requires additional programming expertise and are not available in widely used Bayesian software (e.g., JAGS, Stan, etc.).
This may hinder uptake and use of our proposed model. 
Therefore, we rely on the results and implementation provided by \citet{Shaby2012} which makes implementing our model as accessible to agricultural researchers (and statisticians) as possible. 
A thorough and detailed description of all models presented in this study, and each MCMC algorithm to fit said models, can be found in Appendix A. 
In what follows, we focus on evaluating the utility and reliability of our model using simulated data. 
We then demonstrate our model using data generated by designed agricultural experiments.

\section{Simulation Study}
We conducted a simulation study to evaluate the utility and reliability of our model under scenarios we have encountered in our applied experience. 
We simulated binary data in a spatial domain $\mathcal{S} = [0,2] \times [0,1]$ that was divided into a grid of 18 rows and 32 columns of equally-sized cells.
Each cell represents a plot in a designed agricultural experiment that results in a single recorded observation (i.e., $i = 1, 2, ..., 576$).
The geographical location of each plot is described by the coordinates of the plot's centroid. 
This grid mimics the spatial structure of a designed agricultural experiment. 
We then generated a binary observation for each plot by drawing from a Bernoulli distribution (i.e., $y_i \sim \text{Bernoulli}(p_i)$), with expected value $p_i$. 
We generated $p_i$ using (\ref{eq:2}) with a probit link function and $\mathbf{x}_i$, $\boldsymbol{\beta}$, and $\eta_i$.
We generated $\mathbf{x}_i$ in the same fashion as an RCBD, randomly assigning the treatments to each plot. 
We defined the vector of treatment effects $\boldsymbol{\beta} \equiv (\beta_0 = -0.3,\beta_1 = 0.5)$. 
Last, we generated $\eta_i$ simulating four scenarios with different spatial patterns (see paragraph below for a detailed description of each scenario).
We chose to simulate binary data for all scenarios described below because binary data are common in designed agricultural experiments, appear in our data examples, and perhaps represent the most difficult type of data to estimate a spatial effect from. 
All scenarios, described below, were simulated 1000 times.

Regarding the block effect (or spatial effect), we simulated four scenarios with different patterns for $\eta_i$ in (\ref{eq:2}) (Table \ref{table1}). 
The first scenario describes a block effect in an RCBD with four identical blocks with dimensions $\mathcal{B} = [0,0.5] \times [0,1]$ (see Figure \ref{figure1}).
In our applied agricultural research experience, this scenario represents larger contrived blocks that are forced due to practical constraints for which the statistician has little input.
However, in the simulated data generated by our first scenario, both assumptions mentioned above are met: 1) the spatial blocks can be clearly delineated within a field; and 2) the spatial blocks represent groups of similar experimental units. 
Thus, we named this first scenario ``block assumptions met".
When the block assumptions are met, scenario 1 demonstrates our hybrid model's utility as a model checking tool. 
The 3 scenarios described below illustrate our hybrid model's utility as an alternative modeling approach for agricultural researchers when assumptions 1 or 2 do not hold.
We named the other three scenarios relative to the first one, envisioning a situation where an agricultural researcher receives a protocol to carry out and experiment with an RCBD (i.e., scenario 1), but the data generating block effect is mis-specified or absent relative to scenario 1.
We named the second scenario ``mis-specified blocks" because the spatial blocks cannot be clearly delineated within a field prior to data collection (i.e., assumption 1 does not hold), however, the spatial blocks contain plots that are similar to each other (i.e., assumption 2 holds). 
In other words, scenario 2 is compatible with traditional experiment design, but where blocks were mis-specified in the original block layout. 
We named the third scenario ``no block effect" because it describes a situation with no apparent spatial effect.
In our applied experience in agricultural research, the third scenario appears commonly in designed agricultural experiments, however, we suspect that this finding may be influenced by mis-specified blocks. 
We named the fourth scenario ``smooth spatial effect" because neither assumption 1 or 2 holds (i.e., the notion of blocks does not exist) but a spatial effect is generated by a multivariate normal distribution with a variance-covariance matrix with elements determined by an exponential covariance function.
Scenario 4 appears in \citet{Brownie1993} and \citet{Stroup1994}. 
We fit our hybrid model to the simulated data presented in Table \ref{table1} for evaluating the reliability and utility. 
Again, our goal is to expand the modeling options available to agricultural researchers rather than a comparison of our hybrid model to existing approaches. 

In the spirit of \citet{Little2006}, we evaluated the Frequentist properties of our Bayesian model to assess its reliability to obtain statistical inference. 
We computed the bias and empirical coverage probability of 1000 simulations (i.e., 1000 different simulated datasets).
For our evaluations, we used the contrast of interest $\beta_2 - \beta_1$. 
We used $\text{E}(\beta_2 - \beta_1 | \mathbf{y})$ as a point estimate to evaluate bias as the difference between the true expected value, $\text{E}(\beta_2 - \beta_1 | \mathbf{y}) - (\beta_2 - \beta_1)$. 
Additionally, we obtained the 95\% credible intervals of said contrasts to evaluate the empirical coverage. 
Further details on the model and MCMC algorithm can be found in Appendix A, and further details on the simulation study can be found in Appendix B. 

\begin{table}[h]
    \begin{center}%
    \begin{tabular}{p{0.35\textwidth}p{0.55\textwidth}}
        \hline
        \textbf{Scenario} & \textbf{Data generating spatial effect} \\
        \hline
        1. Block assumptions met & Block effect with 4 blocks\\
        2. Mis-specified blocks & Block effect with 8 blocks \\
        3. No block effect & No spatial dependence pattern\\
        4. Smooth spatial effect & Exponential covariance function\\
        \hline
    \end{tabular}
   \end{center}%
   \caption{Summary of simulated data.}
\label{table1}
\end{table}

\subsection{Results}
\label{makereference3.1}
\begin{figure}[t]
    \includegraphics[width=\textwidth]{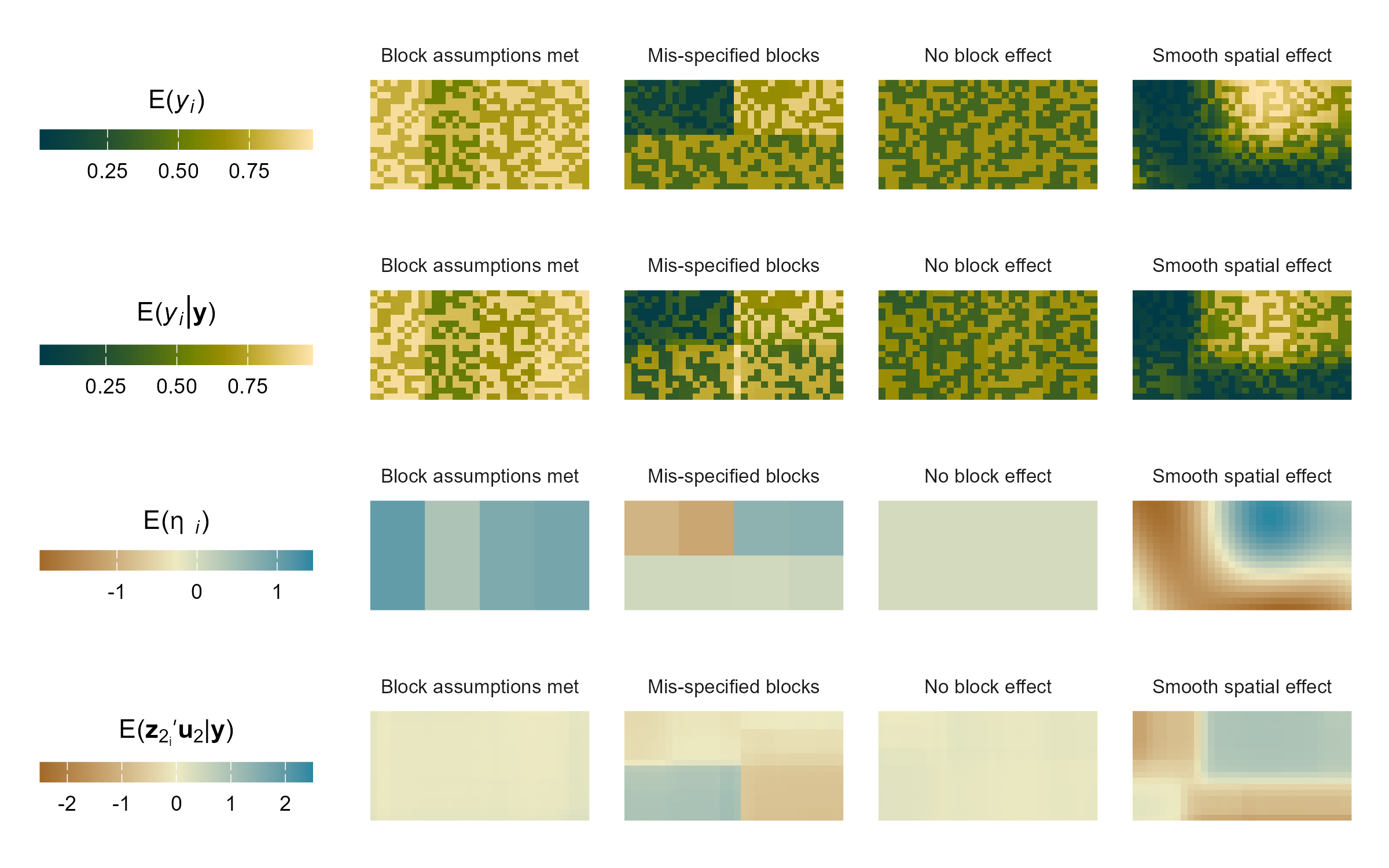}
    \caption[Simulated data]{Simulated data from four scenarios that show the reliability and utility of our hybrid model. 
    The first row and the third row show the true expected value of the response (i.e., $\text{E}(y_i)$) and the true spatial effect (i.e., the sum of block-like effects with different spatial resolutions $\eta_i = \sum_{r=1}^{2} \mathbf{z}_{r_i}' \mathbf{u}_r$) for all four scenarios.    
    The second row and the fourth row show the expected value of the posterior predictive distribution (i.e., $\text{E}(y_i|\mathbf{y})$) and the expected value of the posterior distribution of the estimated block effect (i.e., the second element of the multi-resolution block effect, $\text{E}({\mathbf{z}_{2_i}'}\mathbf{u}_2|\mathbf{y})$) respectively.
    The four scenarios are summarized in Table \ref{table1} and described in the Methods section.
    }
    \label{figure1}
\end{figure}

Our simulation's results demonstrate our model's utility as a visual model checking tool or as an alternative modeling approach for agricultural researchers.
First, analyses of individual simulations demonstrate our hybrid model’s utility as a visual model checking tool. 
To interpret Figure \ref{figure1}, recall that the model includes the designed blocks in ${\mathbf{z}_{1_i}'}\mathbf{u}_1$ and any spatial pattern besides the designed blocks may be recovered by the regression tree estimating ${\mathbf{z}_{2_i}'}\mathbf{u}_2$.
If no such pattern is distinguishable, it may be interpreted as a suggestion that the designed blocks correctly capture the field's spatial processes. 
When the designed blocks correctly capture the field's spatial processes, as in scenario 1, our hybrid model visually appears to estimate no additional spatial effects. 
That is, the second level of the multi-resolution spatial component visually appears to be zero (Figure \ref{figure1}). 
In scenario 2, our hybrid model visually appears to estimate smaller-scale spatial blocks, indicating that the original designed blocks are mis-specified. 
When no block effect is present, as in scenario 3, our hybrid model visually appears to estimate that no spatial effect resembling blocks is present. 
In the fourth scenario, where no block effect is present but a traditional spatial effect is present, our hybrid model visually appears to capture large-scale effects, but fails to capture finer-scale patterns.
The second scenario demonstrates our hybrid model’s utility as an alternative modeling approach for agricultural researchers.
The fourth scenario demonstrates limitations of our hybrid model when the spatial effect follows a correlated multivariate normal distribution and thus, as expected, should be modeled using the approaches presented in \citet{Brownie1997} and \citet{Stroup1994}.

\begin{figure}[t]
    \includegraphics[width=\textwidth]{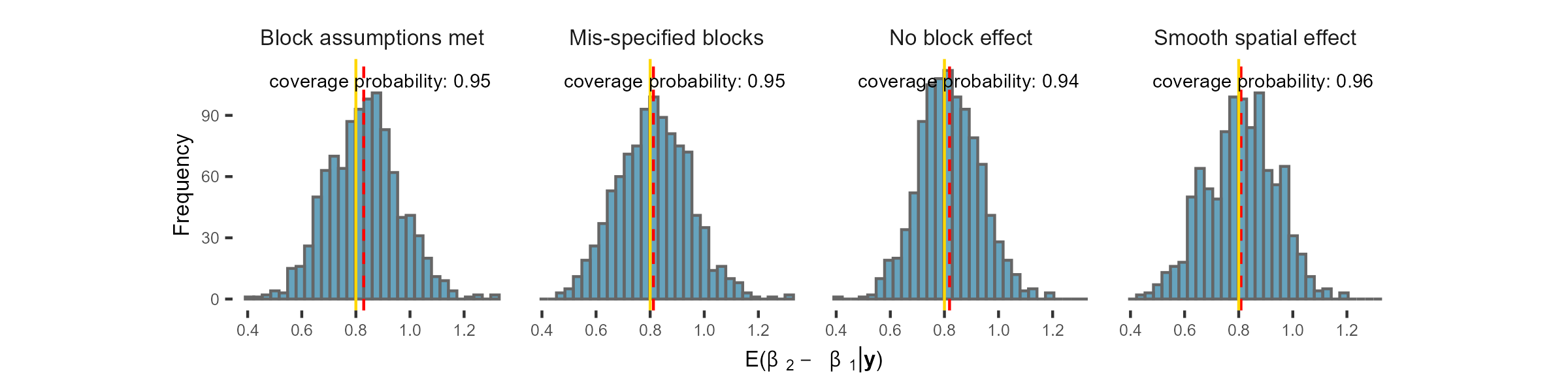}
    \caption[Simulation study]{Results from 1000 simulated data sets mimicking four typical scenarios we have encountered in our applied experience in agricultural research.
    The panels show the expected value of the posterior distribution of the contrast between the first and second elements of $\boldsymbol{\beta}$ (i.e., $\text{E}(\beta_2-\beta_1|\mathbf{y})$) for the four simulated scenarios. 
    The vertical golden line shows the true expected value of the contrast (i.e., $\text{E}(\beta_2-\beta_1)$).
    The vertical red line shows the mean value of all simulations (i.e., $\text{E}(\beta_2-\beta_1 | \mathbf{y})$).
    The text indicates the empirical coverage probability of the 95\% credible intervals.
    The four scenarios are summarized in Table \ref{table1} and described in the Methods section.
    }
    \label{figure1.5}
\end{figure}

In addition, the results of the simulation study support our hybrid model’s utility as an alternative modeling approach for agricultural researchers.
Our model provided reliable results for estimation of the contrast of interest, $\beta_2-\beta_1$, as demonstrated by the Frequentist properties.
The expected values of the contrasts were centered around the true value and their associated 95\% credible intervals were correctly calibrated around 95\% (Figure \ref{figure1.5}, Table \ref{table2}). 
The width of the 95\% credible intervals was only slightly reduced for scenarios without apparent spatial effects (i.e., when $\boldsymbol{\eta}=\boldsymbol{0}$. 
Across all four scenarios, regardless of the underlying spatial pattern, the results were consistent in terms of bias and coverage. 

\begin{table}[h]
    \begin{center}%
    \begin{tabular}{p{0.35\textwidth}cc}
    \hline
    \textbf{Scenario} & \textbf{Coverage probability} & \textbf{$\text{CI}_{95\%}$ width}\\
        1. Block assumptions met & 0.95 & 0.57\\
        2. Mis-specified blocks & 0.95 & 0.56\\
        3. No block effect & 0.94 & 0.48\\
        4. Smooth spatial effect & 0.96 & 0.58\\
        \hline
    \end{tabular}
   \end{center}%
   \caption{Results of simulation study. Coverage probability and 95\% credible interval width of the contrast $\beta_2-\beta_1$. Recall that in the simulation $\beta_2-\beta_1 = 0.8$. }
\label{table2}
\end{table}

\section{Data example I: Wheat resistance to Hessian fly damage}
\label{makereference2.2}

We demonstrate our hybrid model using data from a designed agricultural experiment that was collected to compare the resistance of wheat (\textit{Triticum aestivum} L.) genotypes to damage from Hessian fly (\textit{Mayetiola destructor} Say) \citep{Gotway1997}.
Thus, the experiment consists of a one-way treatment structure in a randomized complete block design. 
The treatment is the wheat genotype for each plot. 
The experiment was conducted in an agricultural field that was divided into a grid of 8 rows by 8 columns, with each plot  $3.7 \times 3.7$ m. 
The blocks were delineated by an agricultural researcher prior to data collection by dividing the field into four quadrants (Figure \ref{figure3}).
At each plot, an agricultural researcher sampled $m_i$ plants and recorded the number of plants with Hessian fly damage ($y_i$).
The number of plants sampled at each plot ($m_i$) varied between 6 and 20. 
Following \citet{Gotway1997}, we assume a Binomial distribution in (\ref{eq:1}).
Similar to our simulated data, we assume a probit link function in (\ref{eq:2}) to implement our hybrid model. 
Further details on the model and MCMC algorithm can be found in Appendix A. 

\subsection{Results}
\label{makereference3.2}

\begin{figure}[h]
    \centering
    \begin{subfigure}{0.5\textwidth}
    \centering
    \includegraphics[width=\textwidth]{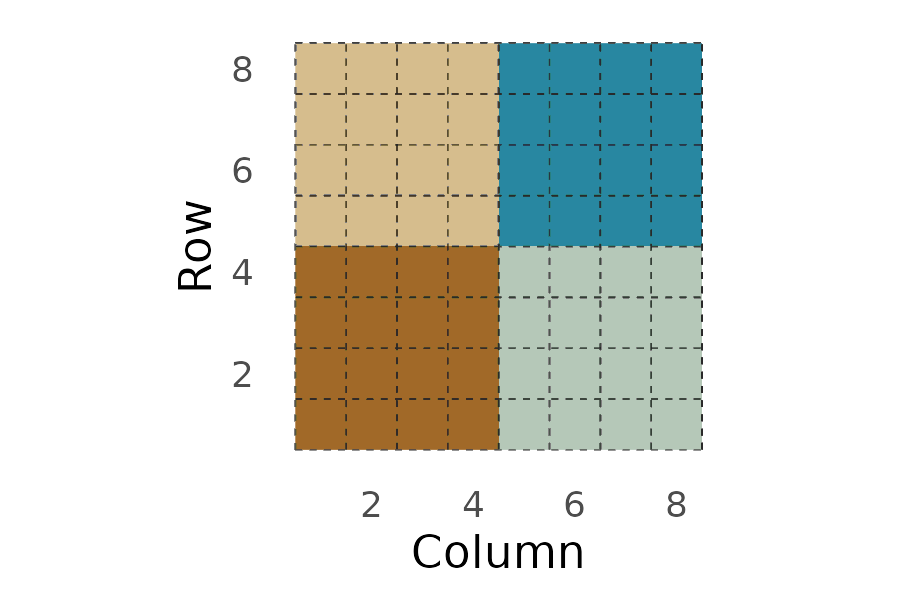}
    \label{fig2a}%
    \end{subfigure}%
    \begin{subfigure}{0.5\textwidth}%
    \centering
    \includegraphics[width=\textwidth]{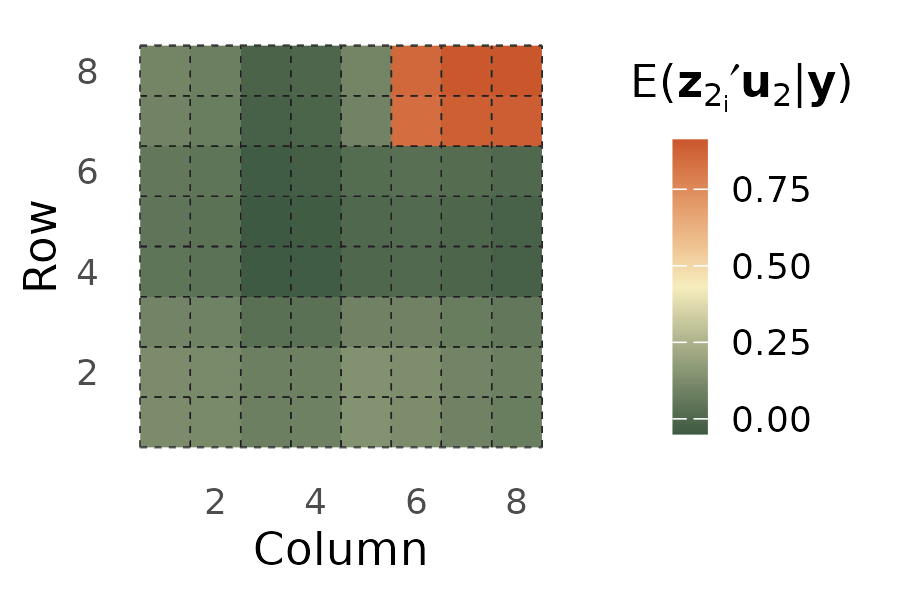}
    \label{fig2b}
    \end{subfigure}
    \caption[Data example I]{The left panel shows the original assumed blocks from a designed agricultural experiment studying wheat resistance to Hessian fly (\textit{Mayetiola destructor} Say) damage taken from \citet{Gotway1997}.
    Each colored quadrant in the left panel indicates an original assumed block.  
    The squares delineated by dashed lines in the left panel show the plots where each observation was recorded. 
    The right panel shows the expected value of the posterior distribution of the second component of the multi-resolution spatial effect ${\mathbf{z}_{2_i}'}\mathbf{u}_2$ (i.e., the spatial effect excluding the block effect, $\text{E}({\mathbf{z}_{2_i}'}\mathbf{u}_2|\mathbf{y})$). 
    Note that our hybrid model estimated additional spatial patterns that were not represented by the original block layout. 
    }
    \label{figure3}
\end{figure}

Our hybrid model estimated block-like spatial effects in a different resolution than the designed blocks of the experiment testing wheat resistance to Hessian fly damage (Figure \ref{figure3}). 
The estimated $\mathbf{z}_{2_i}' \mathbf{u}_2$ show roughly nine irregularly-sized and irregularly-shaped rectangles that complement the designed blocks in a different resolution to account for the spatial processes in the field. 
The 95\% credible intervals for all wheat genotype effects (i.e., all elements in $\boldsymbol{\beta}$) included the point estimates of the model with spatial correction in \citet{Gotway1997} (see Figure S1 in Appendix C).

\section{Data example II. Corn yield trial}
\label{makereference2.3}
We demonstrate our hybrid model using data from a designed agricultural experiment that was collected to compare yields of different corn (\textit{Zea mays} L.) genotypes \citep{Brownie1997}.
Thus, the experiment consists of a one-way treatment structure in a balanced incomplete block design. 
The treatment effect is the corn genotype for each plot. 
The experiment was conducted in an agricultural field that was divided into a grid of 18 rows by 11 columns. 
The blocks were delineated by an agricultural researcher prior to data collection by dividing the field into 18 incomplete blocks (Figure \ref{figure4}).
At each plot, an agricultural researcher harvested all plants and recorded the grain yield (i.e., $y_i$).
Following \citet{Brownie1997}, we assume the data follow a normal distribution in (\ref{eq:1}) with an identity link function in (\ref{eq:2}).
Further details on the model and MCMC algorithm can be found in Appendix A. 

\subsection{Results}
\label{makereference3.3}

\begin{figure}[h]
    \centering
    \begin{subfigure}{0.5\textwidth}
    \centering
    \includegraphics[width=\textwidth]{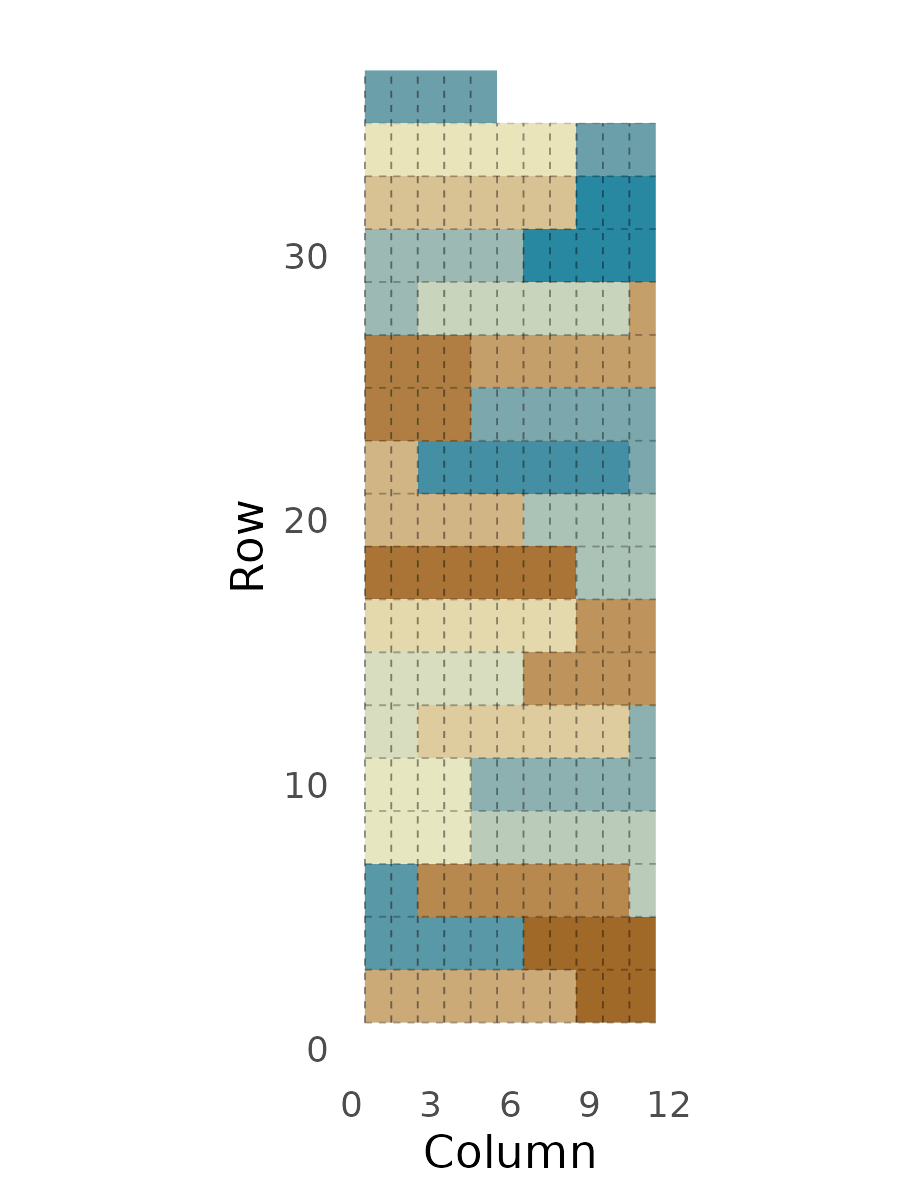}
    \label{fig3a}%
    \end{subfigure}%
    \begin{subfigure}{0.5\textwidth}%
    \centering
    \includegraphics[width=\textwidth]{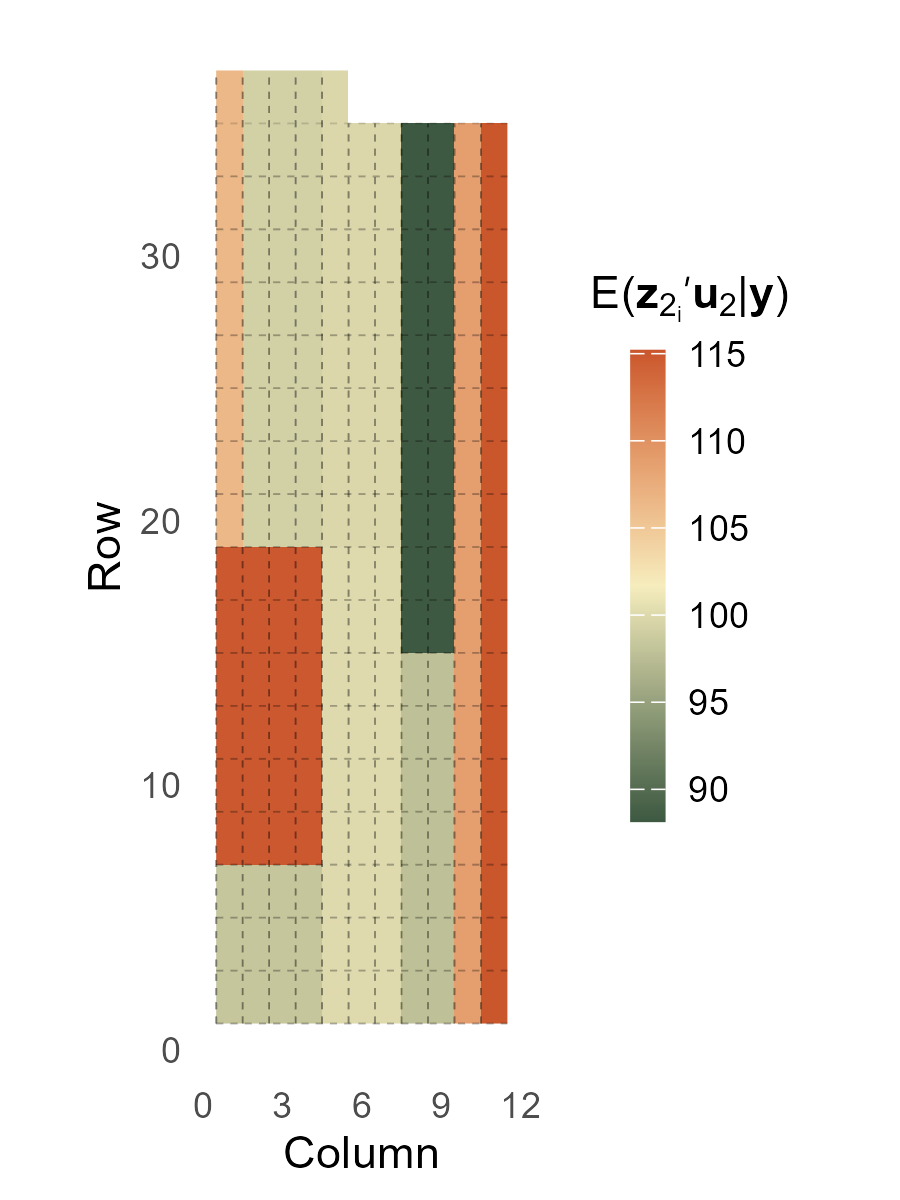}
    \label{fig3b}
    \end{subfigure}
    \caption[Data example II]{The left panel shows the original assumed blocks from a designed agricultural experiment comparing yields of different corn (\textit{Zea mays} L.) genotypes taken from  \citet{Brownie1997}.
    Each colored shape in the left panel indicates an original assumed block.  
    The rectangles delineated by dashed lines in the left panel show the plots where each yield was recorded. 
    The right panel shows the expected value of the posterior distribution of the second component of the multi-resolution spatial effect ${\mathbf{z}_{2_i}'}\mathbf{u}_2$ (i.e., the spatial effect excluding the block effect, $\text{E}({\mathbf{z}_{2_i}'}\mathbf{u}_2|\mathbf{y})$). Note that our hybrid model estimated additional spatial patterns that were not represented by the original block layout. }
    \label{figure4}
\end{figure}

Our hybrid model estimated block-like spatial effects in a different resolution than the designed blocks of the yield trial (Figure \ref{figure4}). 
The original design included 24 blocks, mostly in a horizontal (i.e., east-west) orientation. 
Our results estimated approximately 3 irregularly-sized and vertical (i.e., north-south) larger blocks that complement the designed blocks in a different resolution to account for the spatial processes in the field. 
In addition, two smaller blocks at the bottom divide the field into two asymmetrically-sized areas with different spatial effects. 
The magnitude of the estimated block effect was relatively large, being approximately as large as the treatment effect.

\section{Discussion}
\label{makereference4}
We provide a hybrid modeling framework that combines traditional modeling techniques for designed agricultural experiments (e.g., the GLM framework) with modern statistical techniques.
Our hybrid model is a new alternative for modeling spatial processes using a multi-resolution spatial component that contains blocks with two different spatial resolutions. 
One resolution of blocks corresponds to the designed blocks while the other resolution contains block-like spatial effects estimated with a tree-generating process.
Our hybrid model is based on the widely used GLM framework, combined with the ideas in \cite{Shaby2012}, and serves the dual purpose of validating the originally designed block layout, and estimating the spatial block layout when the original was mis-specified. 
Moreover, understanding the historical trends in the analysis of designed experiments is helpful to put our new model in context: the foundational concepts for modeling data generated by designed experiments in agriculture were established early in the 20th century. 
Some additional features have been integrated steadily down the road \citep{Verdooren2020} and many efforts have been devoted to modeling the block effects ($\boldsymbol{\eta}$) in (\ref{eq:2}). 
For example, some of the most common approaches model $\boldsymbol{\eta}$ as either fixed or random effects, and this choice is at the heart of many of the most contentious debates in the field of experiment design \citep{Dixon2017}. 
Further variations to modeling $\boldsymbol{\eta}$ can be conceptualized as including both spatial blocks and smooth spatial effects (\citealt{StroupGLMM2024}, chapter 18). 
Most importantly, said variations to modeling $\boldsymbol{\eta}$ were incorporated into agricultural experiment analyses borrowing methods from spatial statistics like nearest neighbors analyses \citep{Bartlett1978, papadakis1937methode, Wilkinson1983}, random spatial effects \citep{RICHTER2015}, or polynomial functions \citep{Federer1954}. 
The novelty of our variation for modeling $\boldsymbol{\eta}$ is that it combines the flexibility of modern statistical modeling techniques (i.e., regression trees) with traditional interpretable components (e.g., treatment effects). 
Hybrid models that blend the flexibility of modern techniques with traditional interpretable (e.g., regression and ANOVA) techniques have shown great potential for advancing scientific discoveries \citep{Karpatne_2017,McCormick2021}⁠. 
Moreover, the tree generating process is compatible with blocking concepts and thus, with the traditional designed experiment framework in agriculture. 
Such compatibility of our model with the original designed experiment framework in agriculture may enhance its practical applicability. 
Our hybrid model may serve two purposes: validating the originally designed block layout, or estimating the block layout. 
This flexibility provides researchers with a tool for evaluating and analyzing designed agricultural experiments. 
In the remaining section, we will discuss both validation and estimation purposes, its limitations and opportunities, and further research opportunities. 

First, our results show that our hybrid model can be used conservatively to check the original block design. 
Recall that the two assumptions behind blocked experiments are 1) the spatial blocks can be clearly delineated within a field; and 2) the spatial blocks represent groups of similar experimental units (i.e., similar experimental plots). 
Even though land areas with similar characteristics (i.e., blocks) may exist in the field (i.e., assumption 2 holds), mis-specified blocks are common in agricultural experiments \citep{Stroup1994,Gusmao1986,Gotway1997,Brownie1997}. 
\citet{StroupGLMM2024} highlights that one important reason for mis-specified blocks is blocking for practical convenience (e.g., on-farm field logistics) more than characteristics like soil fertility or topography. 
Likewise, we argue that it is worth checking whether the experiment design overlooks spatial patterns, and provide a model that may be used for that purpose. 
Our hybrid model constitutes a simple and straightforward technique to check for the presence of spatial patterns in addition to the designed blocks.  
Based on our applied experience, we envision three frequently observable conclusions: 
(i) after accounting for the designed spatial blocks, no spatial patterns were discovered (i.e., the designed blocks correctly capture the field's spatial processes, scenarios 1 and 3 in Figure \ref{figure1}), 
(ii) after accounting for the designed spatial blocks, undefined, irregular, large-scale spatial patterns were discovered (i.e., the designed blocks do not capture some of the field's spatial processes) and should be combined with a smooth spatial effect in the statistical model (e.g., scenario 4 in Figure \ref{figure1}), 
and (iii) after accounting for the designed spatial blocks, defined block-like spatial patterns were discovered (i.e., the designed blocks do not capture some of the field's spatial processes) and the block layout should be modified the following year (e.g., scenario 2 in Figure \ref{figure1}).
We envision the implementation of the posterior distributions that delineate the blocks as the priors for subsequent experiments in the same field, a possibility enabled by the compatibility of our tree generating process with blocks in traditional experiment design.

Additionally, our hybrid model offers a novel alternative for accounting for spatial variability for cases with non-traditional spatial dependence and is compatible with traditional experiment design. 
While most spatial components in GLMs involve borrowing methods from spatial statistics that assume relatively smooth variations in space (e.g., stationary data) \citep{brownie1992comparison}, our tree-generating process is capable of detecting non-traditional spatial dependence patterns like abrupt changes in space (i.e., mis-specified blocks). 
In our applied experience, the notion of blocks appears to exist in field settings, where the corrective approaches borrowed from spatial statistics (e.g., stationary Gaussian process) do not capture this non-traditional spatial dependence. 
For example, we have encountered scenarios where non-traditional spatial dependence arise because of previous crops or management practices that were applied in a block-like manner. 
Our simulation study results support the validity of the statistical inference that may be obtained from our model used as an estimation tool. 
Our model may be thus applied for the estimation of the treatment effects in scenarios with non-traditional spatial dependence. 
We envision the application of our model for estimation after discovering that the original block layout does not capture some relevant spatial patterns and before adjusting the following experiment's design.
When the block assumptions hold, our hybrid model may be most useful as a model checking tool, followed by inference from the traditional experiment design framework. 
We suspect that our approach will result in less precise estimations in cases where the blocking assumptions hold, especially for the general linear model. 
For example, similar to our claims about precision, \citet{Stroup2002} reported that replacing block effects for spatial covariance can signify a loss of efficiency when block assumptions actually hold. 
Therefore, we suspect that our hybrid model will provide less precise estimates of $\boldsymbol{\beta}$ for field experiments where block assumptions hold. 

We anticipate future research may build upon our work by (i) developing the prior on the tree-generating process in our hybrid model, (ii) investigating the scalability of this approach, and (iii) exploring additional approaches to combining classical experiment design with novel approaches to modeling GLMs. 
We have presented a case that uses a simple off-the-shelf RT to estimate block effects, where prior information regarding important components of the experiment design (e.g., all treatments are present in a single block) has not been incorporated. 
Future research may include: (1) shape constraints on blocks, (2) treatment allocation constraints on blocks, and (3) alternative models at terminal nodes (e.g., treed Gaussian processes). 
We envision these and other constraints will be incorporated by formal elicitation of expert knowledge from subject matter experts (e.g., an Agronomist) by using the prior distribution to its fullest potential. 
Our MCMC algorithms were completed, on average, in less than 2 minutes on a laptop computer, demonstrating that this approach is computationally accessible.
Investigating the scalability of this approach may be relevant for further applications (e.g., for analysis of on-farm experiments). 
Again, the approach presented in this paper is particularly compelling because it combines concepts in classical experiment design with novel approaches to modeling mis-specified spatial blocks. 
Building on this, we envision the adaptation of additional elements from classical experiment design into our framework. 
Specifically, we aim to incorporate core concepts from analysis of variance that are compatible with GLMs. 
Analysis of variance can be viewed as a tool for exploratory data analysis, helping to understand the sources of variability. 
For example, \citet{Gelman2005} divides the predictors into ``batches", each one corresponding to a source of variation, and compares their standard deviations.

\section{Conclusions}
\label{section:conclusion}
We provided a novel approach to check and estimate spatial blocks from data. 
This research is particularly motivated by our applied experience in agricultural research, where we have encountered block-like spatial effects that were not included in the experiment design. 
Said block-like spatial effects often arise due to previous field management applied in a block-like fashion. 
Field experiments that are affected by such non-traditional spatial effect may have mis-specified blocks and may benefit from the approach we present in this paper.
We use a hybrid model that estimates spatial patterns that resemble spatial blocks (i.e., areas in the field with similar plots) using a Bayesian GLM with an embedded RT. 
Overall, our hybrid model can serve one of two purposes: validating the originally designed block layout, or estimating spatial patterns missing in the original spatial block layout. 
Additionally, our model has further applications in informing the design for subsequent experiments thanks to the compatibility of the tree generating process with blocks in traditional experiment design. 
Thus, this model provides researchers with a novel tool for analyzing data generated by designed agricultural experiments. 

\section{Conflict of interest}
The authors declare no conflict of interest.

\section{Data availability}
The datasets generated during and/or analysed during the current study are available in the GitHub repository \href{https://github.com/jlacasa/blocking_with_RT}{https://github.com/jlacasa/blocking\_with\_RT}.

\bibliography{sample}

\end{document}